# Measurement of shower development and its Molière radius with a four-plane LumiCal test set-up

H. Abramowicz[1], A. Abusleme[2], K. Afanaciev[3], Y. Benhammou[1], L. Bortko[4,b], O. Borysov[1], M. Borysova[1,c], I. Bozovic-Jelisavcic[5], G. Chelkov[6], W. Daniluk[7], D. Dannheim[8], K. Elsener[8], M. Firlej[9], E. Firu[10], T. Fiutowski[9], V. Ghenescu[10], M. Gostkin[6], M. Hempel[4,b], H. Henschel[4], M. Idzik[9], A. Ignatenko[3,d], A. Ishikawa[11], S. Kananov[1], O. Karacheban[4,b], W. Klempt[8], S. Kotov[6], J. Kotula[8], D. Kozhevnikov[6], V. Kruchonok[6], B. Krupa[7], Sz. Kulis[8], W. Lange[4], J. Leonard[4], T. Lesiak[7], A. Levy[1,a], I. Levy[1], W. Lohmann[4,b], S. Lukic[5], J. Moron[9], A. Moszczynski[7], A. T. Neagu[10], F.-X. Nuiry[8], M. Pandurovic[5], B. Pawlik[7], T. Preda[10], O. Rosenblat[1], A. Sailer[8], B. Schumm[12], S. Schuwalow[4,e], I. Smiljanic[5], P. Smolyanskiy[6], K. Swientek[9], P. Terlecki[9], U. I. Uggerhoj[13], T. N. Wistisen[13], T. Wojton[7], H. Yamamoto[11], L. Zawiejski[7], I. S. Zgura[10], A. Zhemchugov[6]

[1] Tel Aviv University, Tel Aviv, Israel
[2] Pontificia Universidad Catolica de Chile, Santiago, Chile
[3] NC PHEP, Belarusian State University, Minsk, Belarus
[4] DESY, Zeuthen, Germany
[5] Vinca Institute of Nuclear Sciences, University of Belgrade, Belgrade, Serbia
[6] JINR, Dubna, Russia
[7] IFJ PAN, 31342 Kraków, Poland
[8] CERN, Geneva, Switzerland
[9] Faculty of Physics and Applied Computer Science, AGH University of Science and Technology, Kraków, Poland
[10] ISS, Bucharest, Romania
[11] Tohoku University, Sendai, Japan
[12] University of California, Santa Cruz, USA
[13] Aarhus University, Aarhus, Denmark

**Abstract** A prototype of a luminometer, designed for a future $e^+e^-$ collider detector, and consisting at present of a four-plane module, was tested in the CERN PS accelerator T9 beam. The objective of this beam test was to demonstrate a multi-plane tungsten/silicon operation, to study the development of the electromagnetic shower and to compare it with MC simulations. The Molière radius has been determined to be 24.0 ± 0.6 (stat.) ± 1.5 (syst.) mm using a parametrization of the shower shape. Very good agreement was found between data and a detailed Geant4 simulation.

[a] e-mail: levy@alzt.tau.ac.il
[b] Also at Brandenburg University of Technology, Cottbus, Germany
[c] Visitor from Institute for Nuclear Research NANU (KINR), Kyiv 03680, Ukraine
[d] Now at DESY, Zeuthen, Germany
[e] Also at DESY, Hamburg, Germany

## 1 Introduction

Two compact electromagnetic calorimeters [1] are foreseen in the very forward region of a detector for a future $e^+e^-$ linear collider experiment, the Luminosity Calorimeter (LumiCal) and the Beam Calorimeter (BeamCal). The LumiCal is designed to measure the luminosity with a precision of better than $10^{-3}$ at 500 GeV centre-of-mass energy and $3\times 10^{-3}$ at 1 TeV centre-of-mass energy at the ILC [2], and with a precision of $10^{-2}$ at CLIC [3] up to 3 TeV. The BeamCal will perform a bunch-by-bunch estimate of the luminosity and, supplemented by a pair monitor, assist beam tuning when included in a fast feedback system [4].

LumiCal and BeamCal extend the detector coverage to low polar angles, important e.g. for new particle searches with a missing energy signature [5]. In the ILD detector [6], LumiCal covers polar angles between 31 and 77 mrad and BeamCal, between 5 and 40 mrad. The LumiCal is positioned in a circular hole of the end-cap electromagnetic calorimeter ECAL. The BeamCal is placed just in front of

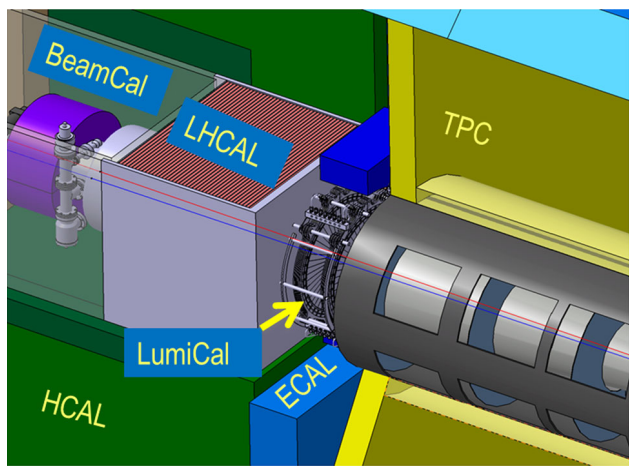

**Fig. 1** The very forward region of the ILD detector. LumiCal, BeamCal and LHCAL are carried by the support tube for the final focusing quadrupole QD0 and the beam-pipe. TPC denotes the central tracking chamber, ECAL the electromagnetic and HCAL the hadron calorimeter

the final focus quadrupole. A sketch of the layout is shown in Fig. 1.

Both calorimeters consist of 3.5 mm-thick tungsten absorber disks, each corresponding to around one radiation length, interspersed with sensor layers. Each sensor layer is segmented radially and azimuthally into pads. The readout rate is driven by the beam-induced background. Due to the high occupancy originating from beamstrahlung and two-photon processes both calorimeters have to be read out after each bunch crossing at the ILC and after a few bunch crossings at CLIC. To ensure a low material budget no cooling infrastructure is foreseen. Hence a dedicated low-power fast readout is developed. Front-end (FE) and ADC ASICs are placed on the outer radius of the calorimeters. In addition, the lower polar-angle region of BeamCal is exposed to a large flux of low energy electrons, resulting in depositions up to one MGy for a total integrated luminosity of 500 fb$^{-1}$ at 500 GeV. Hence, radiation-hard sensors are needed.

The performance of fully instrumented LumiCal and BeamCal detector planes was studied in previous beam-test campaigns. Full functionality of single sensor planes was demonstrated with a signal to noise ratio (SNR) of about 20 for relativistic single electrons [7]. The next step in the detector prototype development was to prepare and conduct a beam-test study of a multi-plane structure, performed in October 2014 at the T9 east area of the proton synchrotron (PS) at CERN. Prototype detector planes assembled with FE and ADC ASICs for LumiCal and for BeamCal have been built. In this paper, results of the performance of a prototype of LumiCal, following tests in the CERN PS beam, are reported.

## 2 Beam-test instrumentation

### 2.1 LumiCal calorimeter prototype

#### 2.1.1 Mechanical structure

To allow the multiple-plane operation, a mechanical structure to meet the demanding geometrical requirements was developed [8]. The required precision of the shower polar angle reconstruction imposes a precision in the positioning of the sensors of a few tens of micrometers. The most important component, the layer positioning structure, includes three aluminum combs with 30 slots each, to install the sensor with the required precision. Since only four detector planes were available to measure the longitudinal shower profile, the mechanical structure had to enable modifications in the prototype layout during the beam test. The overall view of the mechanical structure is presented in Fig. 2.

The LumiCal prototype comprised a mechanical structure of eight tungsten absorber planes interspersed with four fully assembled sensor planes at different positions in between.

#### 2.1.2 Absorber and sensor planes

The 3.5 mm thick tungsten absorber plates are mounted in permaglass frames, as shown in Fig. 3. The inserts located on both sides, right and bottom sides of the permaglass frames, contain small bearing balls providing precisely positioned support-points for the comb slots. For the test presented here, the sensors were mounted onto 2.5 mm thick printed circuit boards (PCB)[1] serving as the mechanical support and high voltage supply. Sensor layers were positioned in slots to form a stack, as shown in Fig. 4 for the first configuration. Different configurations of absorber and sensor planes were used to measure electromagnetic showers at several positions inside the stack.

The positioning precision of the supporting structure, as well as the tungsten absorber thickness uniformity, were extensively tested [8]. The maximum differences between the nominal and measured absorber plane positions do not exceed the required[2] ±50 μm. A function test of the fully assembled stack was performed before installation in the beam test and the full detector prototype functionality was confirmed.

---

[1] This thickness will be reduced in the future to be less than 1 mm using a new connectivity scheme under development. Then sensor planes will be positioned in the 1 mm gap between absorber planes.

[2] This requirement will become relevant when thin sensor planes are used in the future.

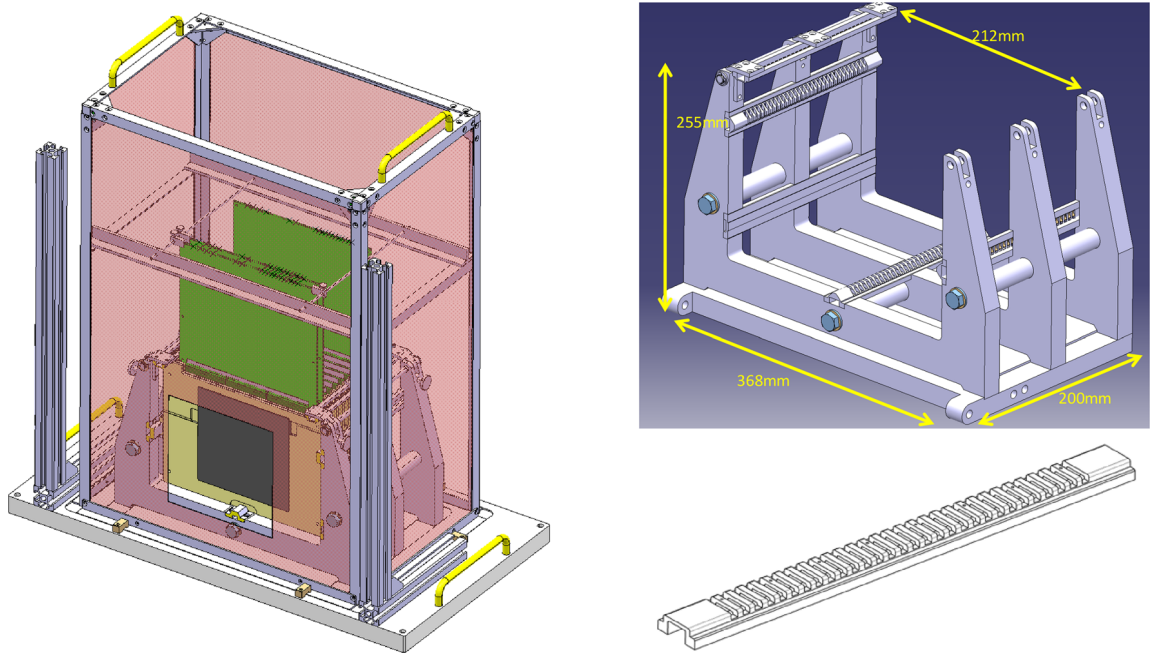

**Fig. 2** Left-hand side: Scheme of the complete mechanical structure with the tungsten absorbers and readout boards installed. Upper right-hand side: Dimensions of the precision mechanical frame for the positioning of sensor and absorber planes. Lower right-hand side: Detail of the retaining comb jig for positioning of sensor and absorber planes

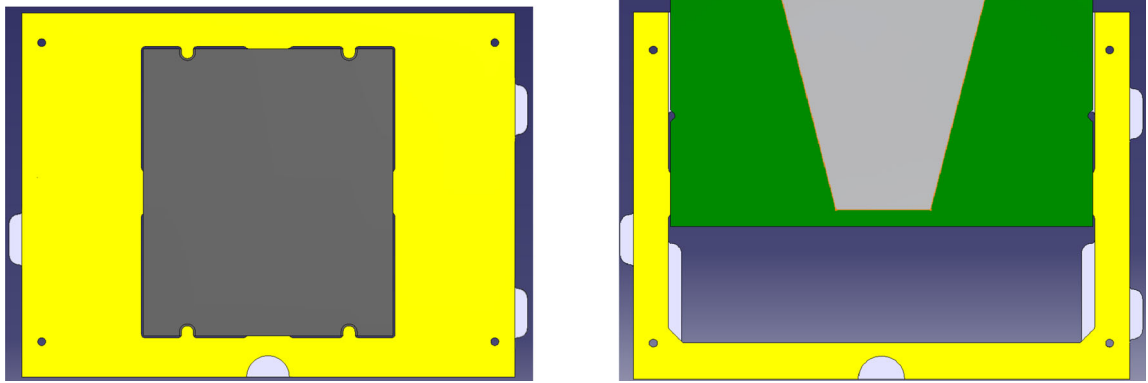

**Fig. 3** Detector layers supporting frames. Left-hand side: tungsten absorber (gray) in a permaglass frame (yellow). Right-hand side: sensor (gray) PCB (green) partially retracted from a permaglass frame (yellow)

### 2.1.3 Detector module

A LumiCal silicon sensor prototype is shown in Fig. 5. It is shaped as a ring segment of 30°, subdivided into four sectors of 7.5° each. The inner radius is 80 mm and the outer radius 195 mm. The thickness of the n-type silicon bulk is 320 µm. The pitch of the $p^+$ pads is 1.8 mm and the gap between the pads 100 µm. Thin printed circuit boards with copper traces are used as fan-outs. Fan-out traces were bonded to the sensor pads through small holes on one end and to the connector to the FE electronics on the other end. For each sensor the pads 51–64 of sector L1 and 47–64 of sector R1 were connected, as illustrated in Fig. 5. The sensors were read out by pairs of dedicated FE and 10-bit ADC ASICs. Each channel of the FE ASIC [9] comprised a charge-sensitive amplifier, a pole-zero cancellation circuit (PZC), and a first order CR–RC shaper. It was designed to work in two modes: physics mode and calibration mode. In the physics mode (low gain),

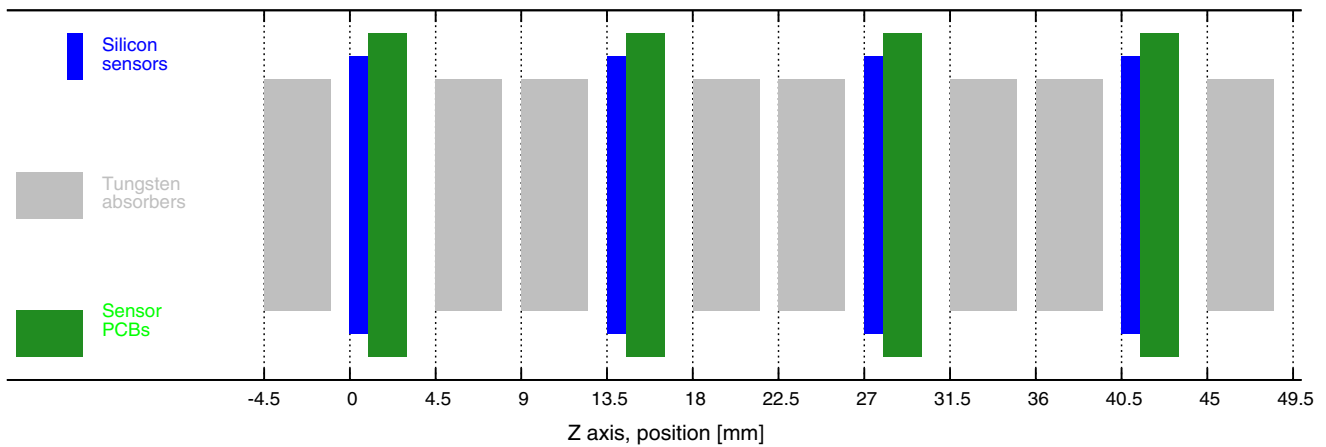

**Fig. 4** Geometry of the first configuration of the LumiCal detector prototype with active sensor layers and tungsten absorbers. The Y axis (perpendicular to Z) is not to scale

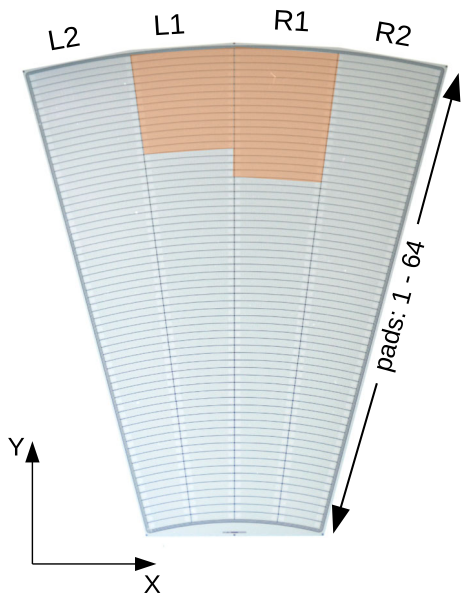

**Fig. 5** A prototype silicon sensor for LumiCal. L1, L2, R1 and R2 are labels for the sectors. The shaded area denotes the pads that were connected to the electronics in this beam test

the dynamic range of the FE is adjusted to accommodate signals typically produced by electromagnetic showers, with up to 10 pC collected charge per channel. In the calibration mode (high gain), the dynamic range covers the signal amplitudes from relativistic muons to be used for calibration and alignment. To match the ILC timing, the shaper peaking time, $T_{peak}$, was set to about 60 ns. The prototype ASICs, containing 8 FE channels, were designed and fabricated in 0.35 µm four-metal two-poly CMOS technology. For historical reasons, in four of the channels an active MOS feedback was used, with about twice the gain of the other four with a passive ohmic feedback, called hereafter R feedback.

To analyze the beam-test data and to perform signal pile-up studies, a sufficiently high ADC sampling-rate and very high internal-data throughput between the ADC and the FPGA-based back-end electronics was assured. An 8-channel 10-bit pipeline ADC ASIC [10] was sampling the signals from the 8 FE ASICs with a rate of 20 MSps (Megasamples per second). The signals from 32 channels (4 pairs of FE and ADC ASICs) were digitized with 10-bit resolution, resulting in a peak data rate of about 6.4 Gb/s. These data were processed by an FPGA-based data concentrator [11]. A photograph of an assembled detector module is shown in Fig. 6.

Upon the arrival of a trigger, 32 channels of each plane were recorded with 32 samplings in the ADC, resulting in a total readout time of 1.6 µs per channel.

### 2.1.4 Calorimeter configurations

For 5 GeV electrons, the expected maximum of the shower is located around the 6th absorber layer. In order to measure the full shower development, at least 10 instrumented layers would be needed. Since only four readout boards were available, this condition would be met only with at least three absorber layers placed between each active sensor layer. However, this leads to a rather small number of measurements which could result in large uncertainties in the shower shape. Since the mechanical support structure enables relatively simple detector geometry changes, a different approach was taken. Three detector configurations were used, with the active sensor layers always separated by two absorber layers. By adding additional absorber layers upstream of the detector, the sensor layers were effectively moved downstream in the shower. The first stack configuration is shown in Fig. 4. A summary of all configurations used is given in Table 1. The single absorber layer after the last silicon sensor was added in order to account for backscattering of particles as expected for the setup in CLIC and ILC detectors. The shower can be therefore sampled up to the 10th layer with a shower sampling resolution of one radiation length. Since the positions of sensors S1–S3 in the first configuration are replicated by the S0–S2 in the second one, the measurement results for the corresponding sensors from both configurations should be almost identical, being a lever arm to control the stability of the response. The data are combined in order to imitate the detector prototype comprising nine active sensor layers. Due to a malfunctioning of the FPGA on the readout board of S3 in the third configuration, only eight positions are used in the analysis.

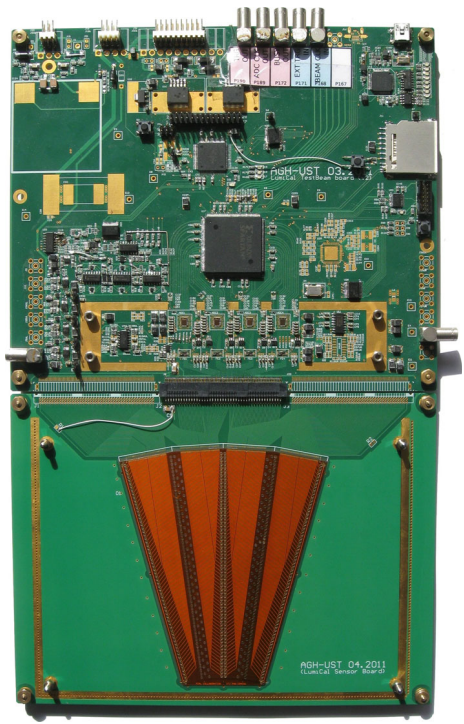

**Fig. 6** Photograph of a LumiCal readout module attached to a sensor

### 2.2 Beam-test set-up

The PS accelerator provides a primary proton beam with momentum of 24 GeV/c. The beam for the T9 area is provided in 400 ms long spills with a typical time separation of 33.6 s between them. The primary beam hits a target, producing the secondary beam to the T9 area which consists of a mixture of electrons, muons and hadrons with momenta in the range of 1–15 GeV/c. A narrow band of particle momenta centered at 5 GeV was selected using a dipole magnetic field and a set of collimators. Electrons and muons were triggered using Cherenkov counters. A pixel telescope [12] was used to measure the trajectories of beam particles. The simplified overall view of the beam and the experimental set-up is presented in Fig. 7.

A schematic diagram of the instrumentation geometry is shown in Fig. 8.

#### 2.2.1 Telescope

To reconstruct the trajectories of beam particles, a four-layer tracking detector, the so-called telescope, was used. The telescope utilizes MIMOSA-26 chips, a monolithic active pixel sensor with fast binary readout [13]. One MIMOSA-26 chip comprises $1152 \times 576$ pixels with 18.4 μm pitch, resulting in an active area of $21.2 \times 10.6$ mm$^2$. The binary readout accepts the pixel signals exceeding a preset discrimination level. The pixel matrix is read continuously providing a complete frame every 115.2 μs. The data are gathered, triggered and stored by a custom DAQ system, based on the National Instrument PXI crate, developed by the Aarhus University in collaboration with the Strasbourg University. The telescope planes, each comprising one MIMOSA-26 chip, were set upstream of the stack as shown in Fig. 8.

Three scintillation counters were used to provide a trigger for particles traversing the active part of the telescope sensors and the region of the sensors in the stack being read out. Two $5 \times 5$ cm$^2$ scintillator tiles were placed upstream and downstream of the telescope and one, with a 9 mm diameter circular hole, was placed just in front of the last telescope plane, as shown in Fig. 8. Photomultipliers were attached to the scintillators. In order to ensure that triggers are only generated by beam particles in the sensitive area of the telescope, the signal from the hole scintillator was set in anti-coincidence.

**Table 1** Positions of active sensor layers in three configurations expressed in number of absorber layers (i.e. radiation lengths $X_0$ of the absorber) in front of the corresponding sensor layer. S0–S3 stands for Sensor 0–Sensor 3

| Configuration | Radiation lengths in number of absorber layers | | | | | | | | | |
|---|---|---|---|---|---|---|---|---|---|---|
| | 1 | 2 | 3 | 4 | 5 | 6 | 7 | 8 | 9 | 10 |
| 1 | S0 | | S1 | | S2 | | S3 | | | |
| 2 | | | S0 | | S1 | | S2 | | S3 | |
| 3 | | | | S0 | | S1 | | S2 | | S3 |

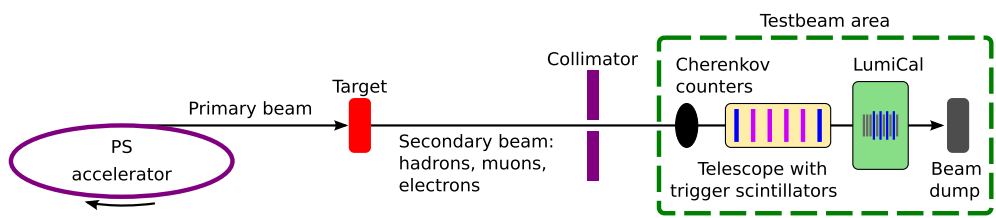

**Fig. 7** Schematic view of the beam and experimental set-up (not to scale)

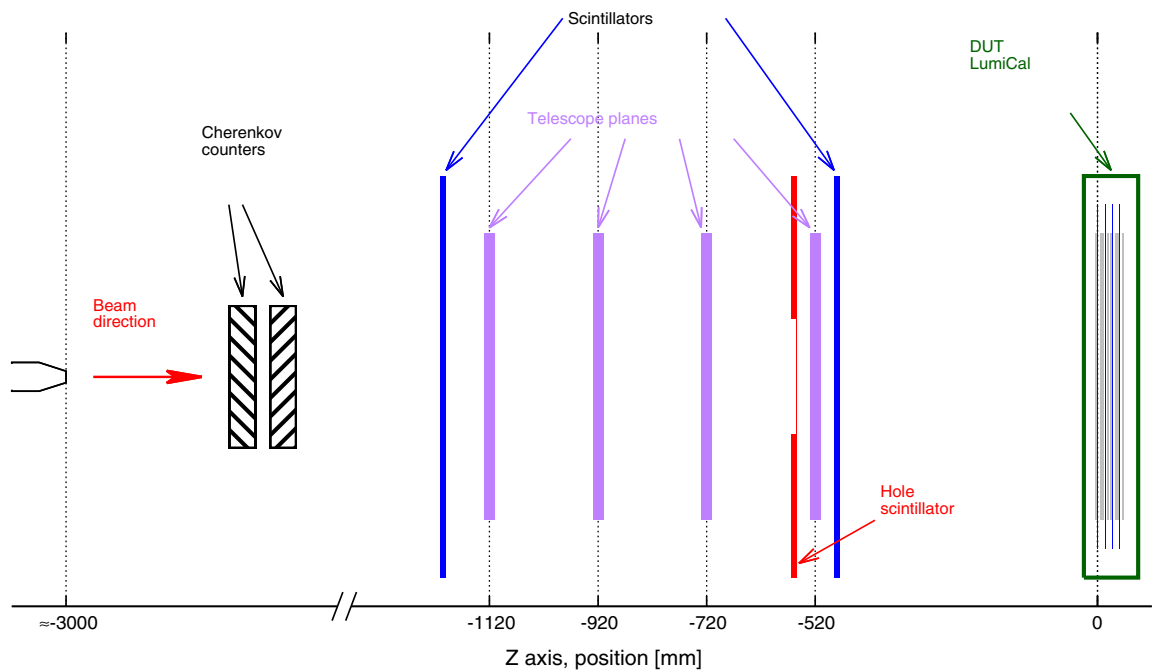

**Fig. 8** Beam test area instrumentation geometry. Not to scale

The trigger signal was combined with the Cherenkov counters response to create a trigger for leptons. A small fraction of particles passing through the active area of the anti-coincidence scintillator is not vetoed due to the inefficiency of the scintillator.

The number of events accumulated is determined by the spill structure of the beam, the rate capability of the telescope and the small electron content in the 5 GeV secondary beam. Of the $10^3$–$10^4$ particles in the beam per 400 ms spill with about 2 spills per minute, only 5% were electrons. The trigger rate in the spill had to be limited, since the MIMOSA-26 chip of the telescope provides continuous readout. Otherwise within the readout time of about 400 µs a second particle may cross the telescope, and the mapping to the electromagnetic shower will become ambiguous. In addition, the DAQ applied a veto against triggers during event data packing, thus rejecting some of the valid electron triggers. Typically, a few electron events per second were registered.

### 2.2.2 Data acquisition (DAQ)

Two independent DAQ computers were used for the telescope and for LumiCal, respectively. Figure 9 displays the connections and the flow of information in the system.

In order to monitor and distribute the trigger signal correctly, a trigger logic unit (TLU) [14,15] was used. The TLU receives the trigger signal and generates an integer TLU number, counting the number of triggers received. The TLU then passes on the trigger signal and the TLU number to the telescope and the LumiCal, respectively. In order to preserve the TLU number and the telescope frame number, a dedicated auxiliary (AUX) unit was used, saving these two numbers for the same event. The AUX unit was also responsible for delivering a long BUSY signal back to the TLU in order to veto triggers during the long readout time of the telescope.

For each event, the TLU number from the LumiCal DAQ, the TLU number from the AUX and the telescope frame num-

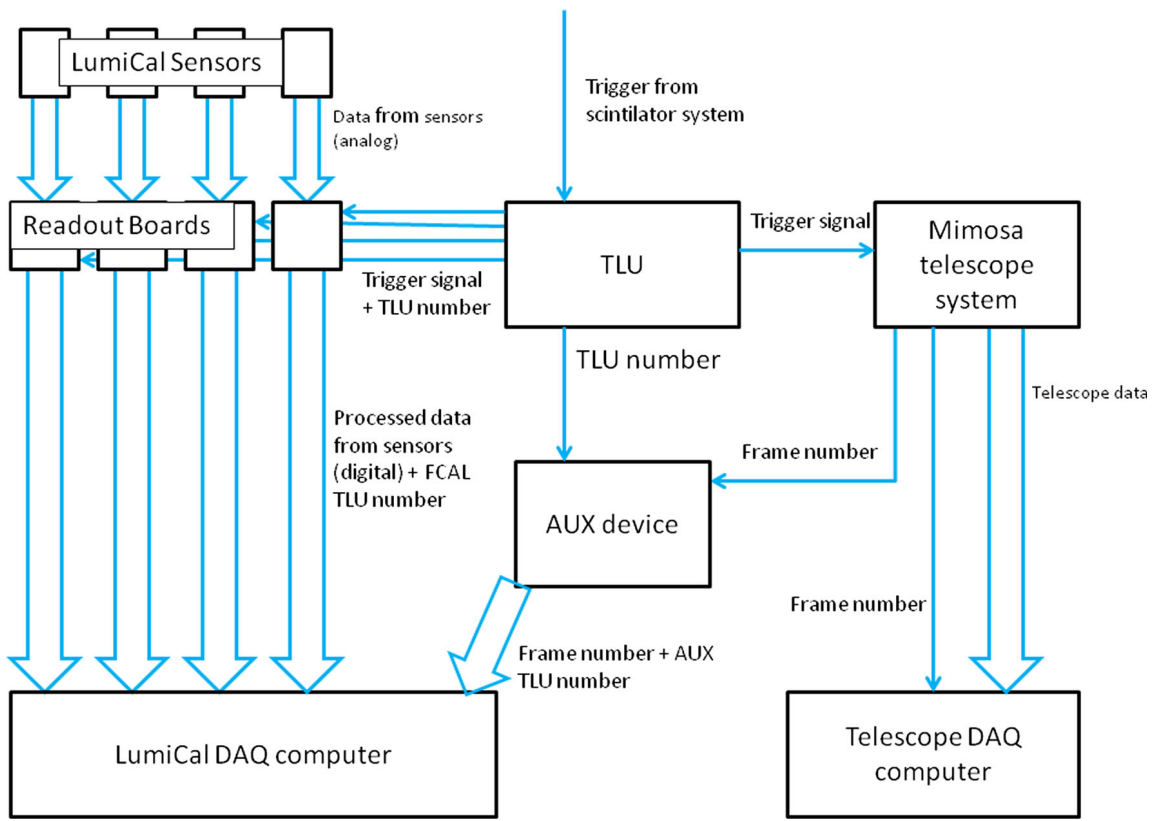

**Fig. 9** A schematic of the connections and the path of the signals in the system. Data is symbolized by a thick arrow, while simple bit information (e.g. the trigger signal) is symbolized by a thin arrow

ber from the AUX were stored to synchronize the LumiCal and the telescope data.

## 3 Data analysis

### 3.1 Telescope alignment and tracking

Since on average for the given trigger and beam intensity a single particle track per event was estimated, the expected number of hits per event in each telescope plane is around one. However, the amount of noise hits dominates over hits generated by the beam particles. The telescope noise is reduced in the analysis by the collinearity requirement between the four telescope planes for hits belonging to a reconstructed track.

As the positions of the telescope planes were only roughly set by the telescope mechanical support structure, an alignment was required to define and correct for the telescope planes relative positions. This was done using the standard procedure implemented in the telescope analysis framework (TAF) [16]. The results can be seen in the hit-maps for the reconstructed tracks, shown in Fig. 10. The beam profile can be clearly seen in the distribution of the reconstructed tracks in the X–Y plane. In particular, the distribution for the last telescope plane remains, as expected from the geometry, in very good agreement with the hole in the anti-coincidence scintillator. From the residuals of the line fit to the track candidates, a hit resolution of 9 μm was determined [17].

The analysis of track angles and associated energy deposits in the LumiCal prototype shows that about 3% of all tracks result from noise hits in the telescope accidentally appearing within collinearity tolerance criteria. These tracks are associated with LumiCal deposits close to zero. In addition, about 2% of all tracks can be attributed either to the inefficiency of the anti-coincidence scintillator or to noise tracks in the telescope associated accidentally to non-zero energy deposits in the prototype.

### 3.2 Data processing in LumiCal detector

The sampling clock of the ADC is running continuously. The trigger for the stochastically arriving beam particles is hence not synchronized with the ADC clock of 20 MHz, leading to an asynchronous sampling. With the rather low expected event rate, 32 ADC samples per event were collected in order to boost the data processing. First, an initial treatment of the data was performed followed by reconstruction of signal amplitudes using a deconvolution procedure [18].

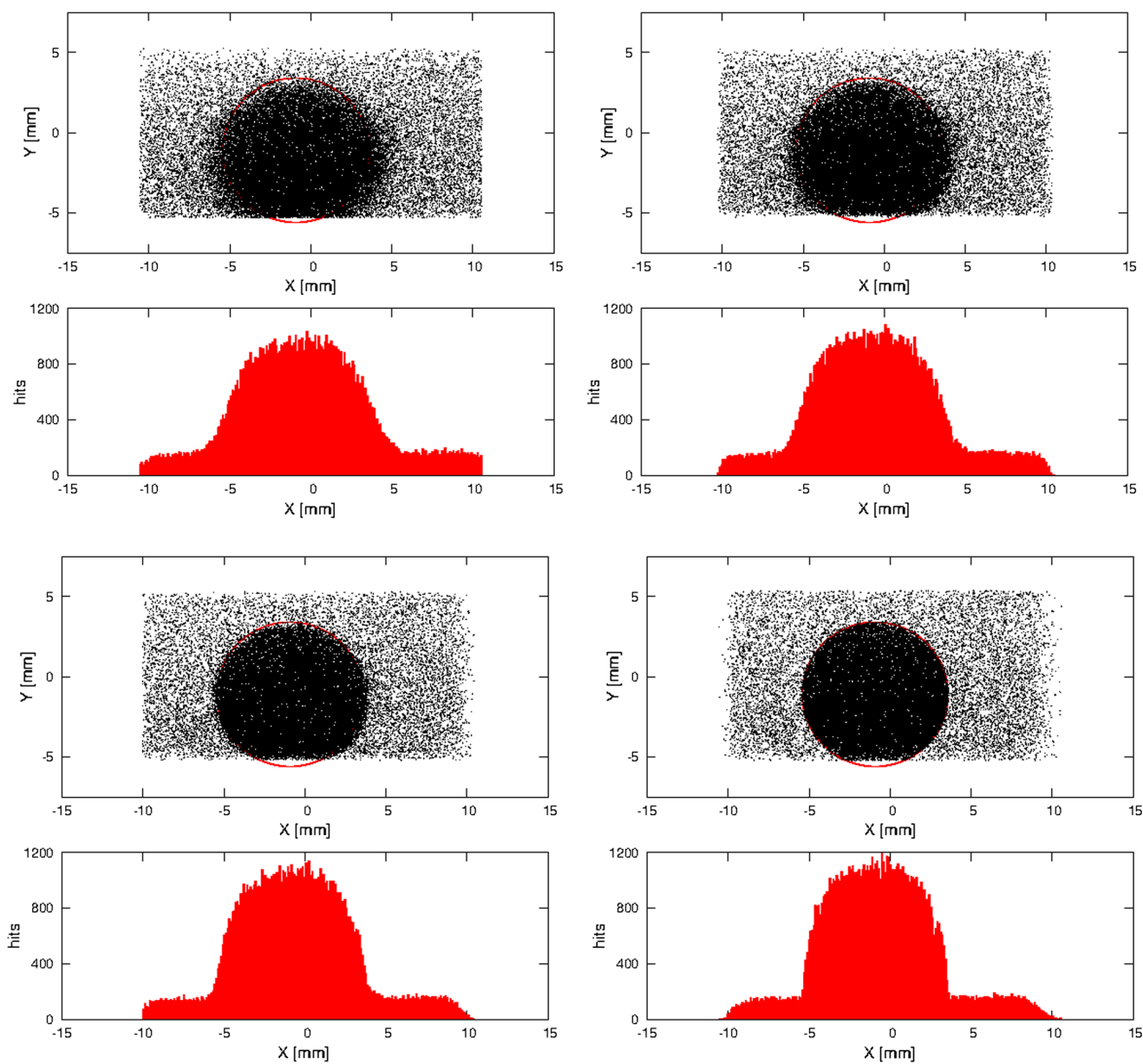

**Fig. 10** Distributions of telescope hits assigned to a track in the X–Y plane after data processing, (top part of figure) together with the projection on the X axis (bottom part of figure). The red circle represents the hole in the anti-coincidence scintillator. The four figures show the hits in each of the four planes, respectively

### 3.2.1 Initial treatment of the data

As can be seen in Fig. 11, the baseline in raw events varies as a function of time synchronously in all channels. This effect is denoted hereafter as common-mode noise presumably caused by power lines. The initial treatment of the data includes the baseline and the common-mode noise subtraction. These procedures have been optimized in order to achieve the lowest readout noise and are described in detail in Ref. [19]. After the baseline and common-mode noise subtraction, the raw event presented in Fig. 11 is transformed to the event shown in Fig. 12.

### 3.2.2 Signal reconstruction by deconvolution

Since the LumiCal readout utilizes an asynchronous ADC sampling, the pulse amplitude is not directly available from the initially processed data and, therefore, has to be reconstructed. The method with the highest expected precision is using a pulse-shape fitting with a pre-defined shape. How-ever, the pulse-shape fitting is a rather slow procedure which requires an initial guess of the fit parameters, and due to the required time can be done only during the offline analysis. In addition, its complexity would be even higher if pile-up of signals occurs.

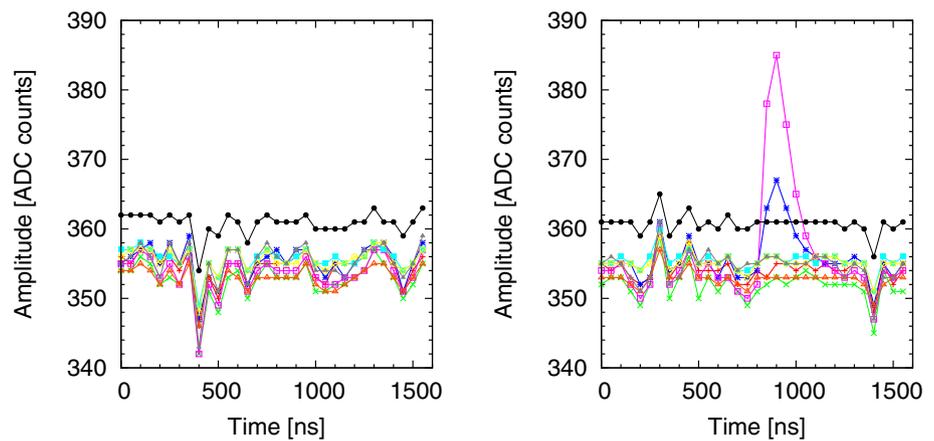

**Fig. 11** Raw amplitudes of two sets of 8 channels, drawn in different colors, as a function of time. One set (left) contains no signal and one (right) shows a signal in two channels (pink and blue)

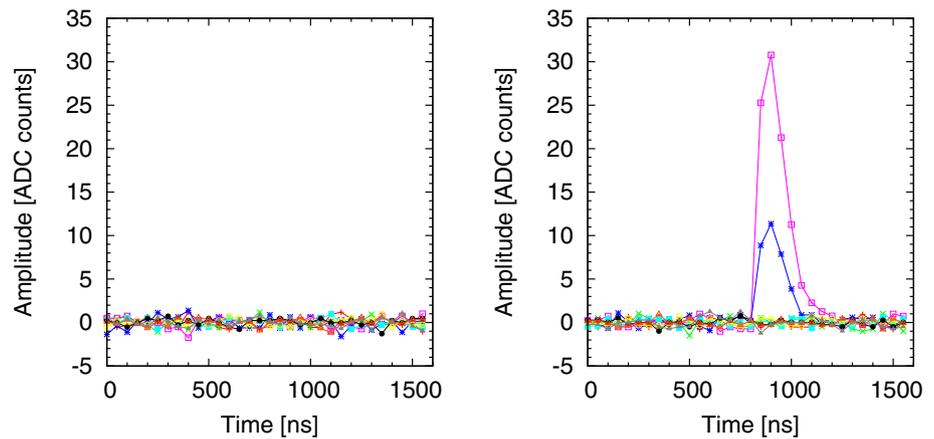

**Fig. 12** The amplitudes presented in Fig. 11 after processing the raw LumiCal data

In order to enable a fast amplitude reconstruction, a deconvolution method [18] is applied for the LumiCal readout. The advantage of this method is the possibility to implement it inside the system digital signal processing block, significantly reducing thus the total amount of data transmitted from the detector. A simple $CR-RC$ shaping was implemented in the LumiCal FE electronics in order to reduce the complexity of the deconvolution filter [19]. The details of the deconvolution have been discussed elsewhere [7,19].

After the initial treatment and deconvolution, applied offline, the signals in the four planes of an electron event in configuration 1 are shown as an example in Fig. 13. The signal amplitudes and number of channels with a signal in neighboring planes show the expected shower development.

### 3.3 Calibration with Muons and signal to noise ratio

The triggered beam particles contained a few percent of muons. Figure 14 shows the raw energy spectrum for a beam containing electrons and muons, obtained as the sum of all 128 channels in the stack for each event. The narrow peak around 120 counts corresponds to the expectations for a muon of 5 GeV and is hence considered as originating from the muons in the beam. It is well separated from the wide distribution expected for electrons. Based on the distribution in Fig. 14, a threshold of 550 ADC counts was chosen to separate the electron events.

The gain of each channel is determined using the muons in the beam. A convolution of a Gaussian and a Landau distribution is fitted to the measured spectra of the deposited energy in each channel. Two examples for such spectra are shown in Fig. 15. As can be seen, for MOS-feedback FE channels the gain is about two times higher and the ratio between the Landau peak position and its width is smaller than for the R-feedback channels.

Using the obtained most probable value (MPV) of the Landau distribution and the noise RMS ($\mathrm{RMS_{NOISE}}$) calculated from baseline fluctuations after deconvolution, the SNR for each readout board in each configuration is calculated as:

$$\mathrm{SNR} = \frac{\mathrm{MPV}}{\mathrm{RMS_{NOISE}}}. \qquad (1)$$

**Fig. 13** The amplitudes as a function of time in the 32 channels of an example electron event after processing the signal in plane 1 (upper left), plane 2 (upper right), plane 3 (lower left) and plane 4 (lower right)

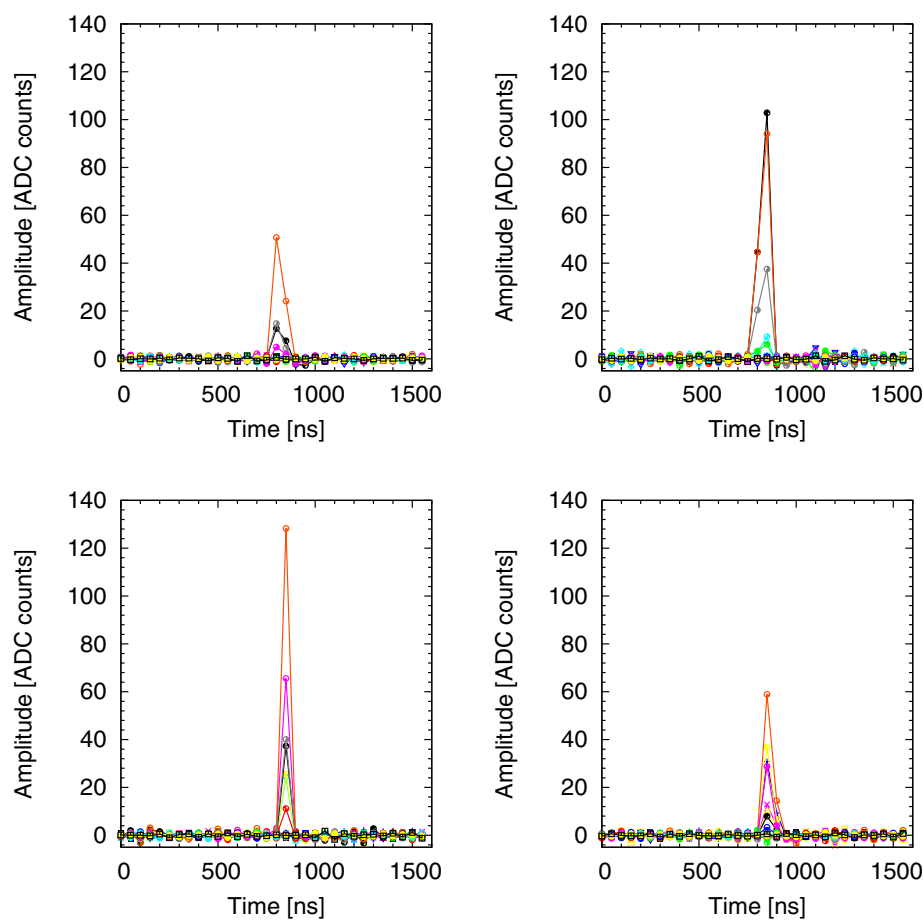

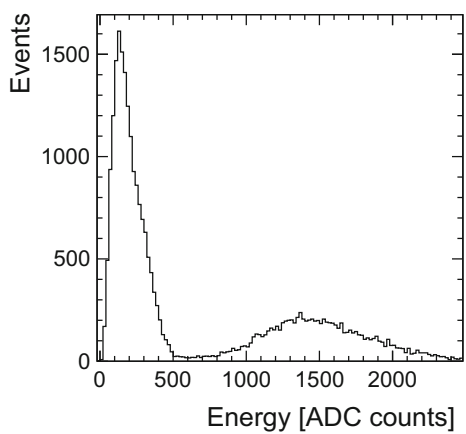

**Fig. 14** Energy deposition in the calorimeter for a beam comprising electrons and muons

The SNR is shown in Fig. 16 for the channels of the four different sensor planes.

The SNR of the R-feedback channels is substantially lower than that of the MOS-feedback ones. This is explained by an increase of the system readout noise in the four-plane detector setup. The MPV depends directly on the channel gain, while the noise introduced by the readout electronics comprises a number of components in the signal processing chain. Since in the past beam-test with a single detector plane the SNR was similar for both types of channels [7], the improvement of noise performance of the multi-plane detector setup will be one of the important future tasks. However, even the lowest SNR value measured in S2 for the R-feedback case is still sufficient for further data processing.[3]

In order to use the same energy scale in both data and simulation, the response for muons was used. The value of the MPV of muon depositions was defined as one unit of minimum ionizing particle (MIP). This value was used to scale all energy measurements.

### 3.4 MC simulation

The beam-test setup was simulated using LUCAS [20], the LumiCal simulation program. LUCAS is a C++ object-oriented toolkit, based only on GEANT4 [21] and ROOT [22]. It was derived from the code written for the ILD detector software package Mokka [23]. In the simulation, the stan-

---

[3] In board S2, the bias currents were not well calibrated, as discovered after the beam test.

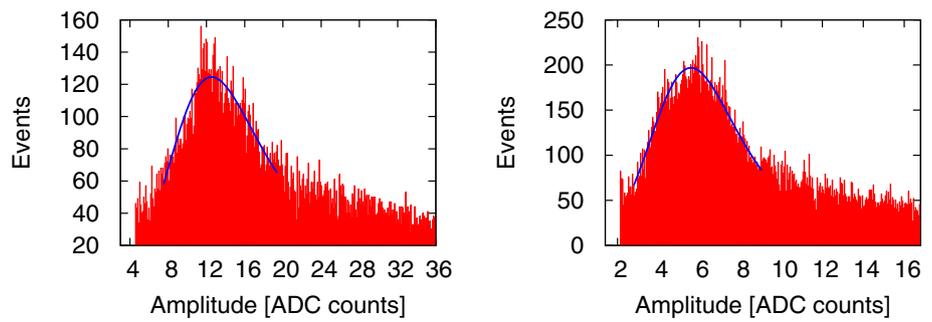

**Fig. 15** Distributions of the signal amplitudes of muons in two readout channels fitted with a convolution of a Gaussian and a Landau distributions. Left: with MOS-feedback, Right: with R-feedback

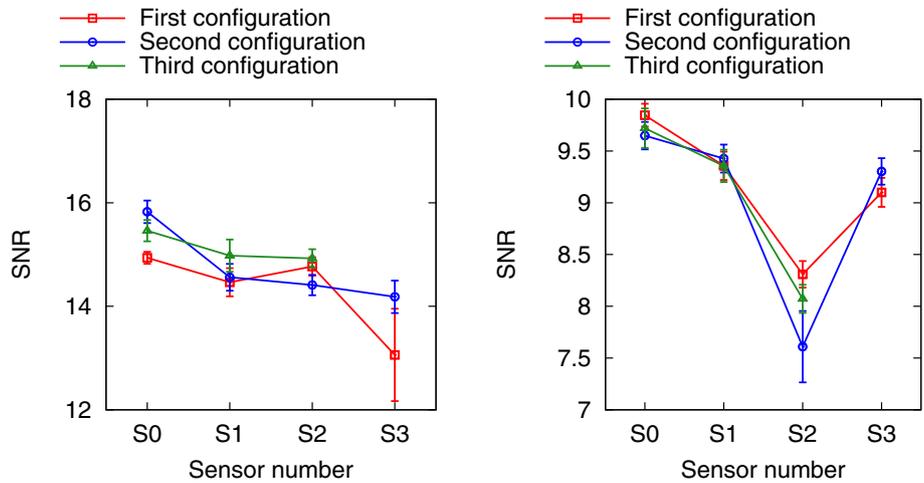

**Fig. 16** The SNR measured in the boards S0 to S3 when placed in configurations 1 (red), 2 (blue) and 3 (black) of the test set-up as explained in Table 1. Left: Channels with MOS-feedback. Right: Channels with R-feedback

dard QGSP_BERT [24] physics list with the GEANT4 range cut-off of 5 µm is used.

The silicon sensor is implemented as a sensitive area of arc shape and the area is distributed to virtual cells. The implementation of the sensor includes the PCB sensor board, with its metalization layer, and the kapton fanout with the copper traces. The beam was simulated as coming from the telescope area, one meter away from the LumiCal stack of square shape with an area of $7.6 \times 7.6$ mm$^2$. The middle of the square was shifted so that the ratio of hits between the two connected sectors will equal the ratio in the data. The extrapolated hit position on the LumiCal front layer from the simulated tracks is presented in Fig. 17. To each hit a color was assigned, corresponding to the pad where energy was deposited. The track angular distribution was generated such that the distributions of the track projection on the X and Y axes reproduce the T9 beam characteristics with an RMS of 0.6 mrad.

## 4 Results

### 4.1 Electromagnetic shower

In order to analyze the longitudinal development of the electromagnetic shower, the distributions of the sum of the

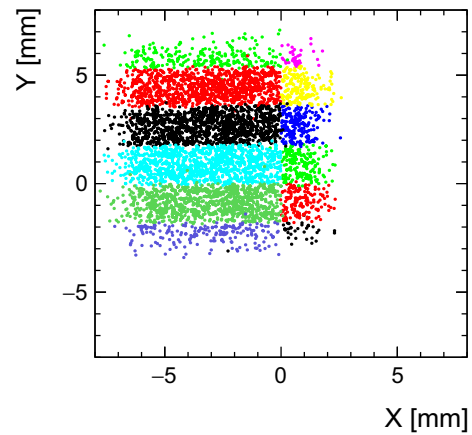

**Fig. 17** In the MC simulation, distribution of the extrapolated hit position in the first layer of LumiCal, with a color code determined by the pad number where the energy was deposited

deposited energy per layer, in MIP units, as shown in Fig. 18, are used. From the first configuration, in which the first sensor layer was placed after one absorber, the energy sum distribution of the first layer is presented in Fig. 18a. From the second configuration where the first sensor layer was placed after three absorber layers, the distribution of the deposited energy in layer 5 is shown in Fig. 18b and for layer 9 in

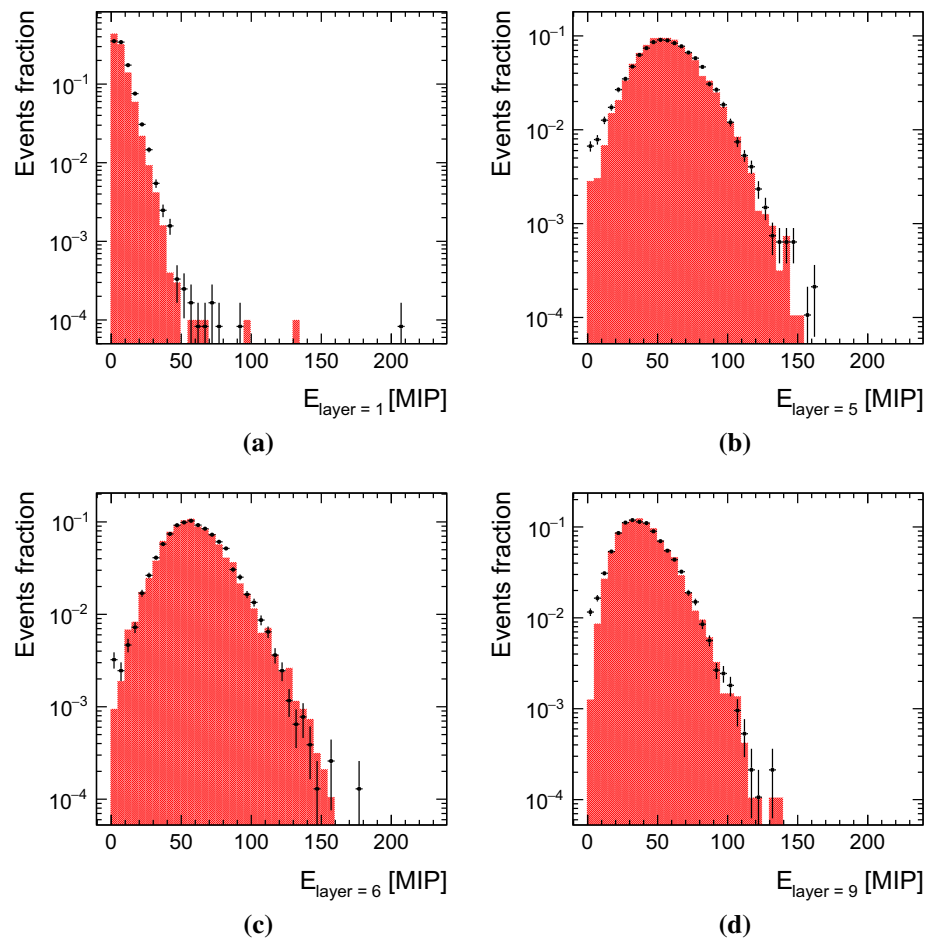

**Fig. 18** Distribution of the energy sum per layer, $E_{layer}$, in MIP units, for layers 1 (**a**) from configuration 1, layers 5 (**b**) and 9 (**d**) from configuration 2 and layer 6 (**c**) from configuration 3. The data (dots) are compared to the MC simulation (shaded region)

Fig. 18d. In Fig. 18c layer 6 is shown, from the third configuration where the first layer was placed after four absorber layers. In all four cases, also the MC simulation is shown and agrees well with the measurements.

The uncertainty on the data is dominated by a 5% calibration uncertainty. This was applied to the average energy measured in six of the nine layers. However, since layers number 3, 5 and 7 were sampled in both configurations 1 and 2, the uncorrelated uncertainties were reduced by $\sqrt{2}$. Systematic uncertainties in the simulation result from the measurement uncertainties of the plate thicknesses which are estimated using local derivatives of the shape, as done in Ref. [25]. They are progressively higher towards the last layers, reaching 2% in the 9th layer.

The longitudinal development of electron showers is shown in Fig. 19a in terms of average shower energy deposits per plane as a function of the number of absorber layers. In Fig. 19b the development according to the three different configurations are presented, each color represents one configuration. Here the common layers in different configurations can be compared and are in good agreement with each other. The results were compared with the prediction of the simulation, and agreement between the simulation and the data is found within the uncertainties (not shown for the simulation). The shower maximum is observed after 6 radiation lengths. The difference in layer 1, where the simulated deposition is slightly smaller, is understood as due to preshowering caused by upstream elements.

### 4.2 Resolution of the position reconstruction

The silicon sensor has a fine segmentation with a pitch of 1.8 mm in its radial direction (pad), as shown in Fig. 5, and, at the location of the beam spot, an arc length in the azimuthal direction of 2.5 cm with a sagitta of 0.4 mm. For further consideration it is convenient to use a Cartesian coordinate system in the transverse plane with the origin coinciding with the shower axis, the Z axis, and the Y axis running along the radial direction of the LumiCal sensor. In this case we neglect the arc shape of the pad by considering it as a strip. Denoting $E_{nkl}$ as the energy deposited in the sensor pad for layer $l$, sector $k$ and radial pad index $n$, the one-dimensional deposited energy distribution for one event along the Y axis can be obtained from the following sum:

**Fig. 19** Average energy deposited in the detector planes of the LumiCal detector prototype as a function of the number of tungsten absorber layers for three configurations combined (**a**) and separately (**b**). The dots (squares and triangles) are data and the shaded area corresponds to the MC simulation

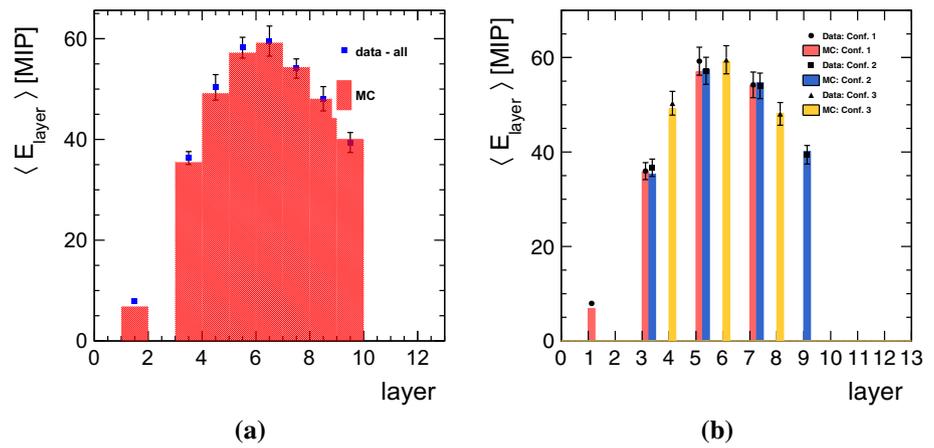

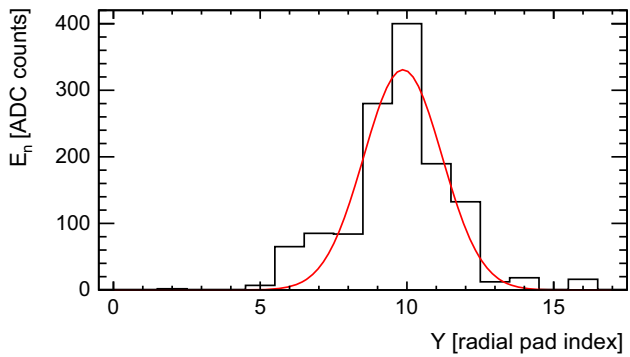

**Fig. 20** The deposited energy sum for a single event, $E_n$, as a function of $Y$ expressed in terms of the pad number. The curve is a Gaussian fit to the data

$$E_n = \sum_{k,l} E_{nkl} , \qquad (2)$$

where the layer index $l$ runs over all sensitive planes in the stack from 1 to 4 and the sector index $k$ for two sectors according to the part of the sensors connected to the readout electronics. The energy of a tower at the pad index $n$ is denoted by $E_n$. An example of the $E_n$ distribution for a single event is shown in Fig. 20. The position of the beginning of the shower on the surface of the first layer projected to the radial coordinate, $Y$, can be estimated as the mean of the Gaussian fitted to the $E_n$ distribution.

The distribution of radial shower position $Y$ is shown in Fig. 21. Most of the estimated shower positions are in the radial pads within the trigger area which spreads over 10 mm. For comparison, the distribution of the extrapolated radial hit position from the beam Telescope track is also presented. Both distributions, which correspond to the beam profile, are in good agreement.

For each event, the reconstructed position of the shower in the radial direction is compared with the extrapolated track impact point position provided by the telescope. The distribution of the difference in the $Y$ coordinate, $\Delta Y$, together with the Gaussian fit, are shown in Fig. 22. Since the telescope position resolution is much better than that of the LumiCal, the resolution of the latter is obtained from the standard deviation of the fit of $505 \pm 10\ \mu$m. This number is the result of multiple scattering and the statistical fluctuations in the shower development, and is in rough agreement with that obtained from the simulation, which gives a resolution of $480 \pm 10$ μm.

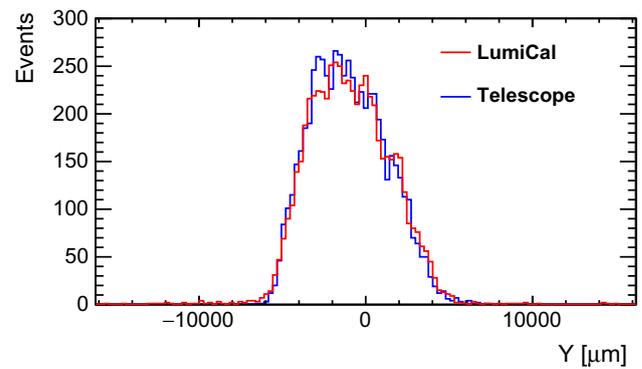

**Fig. 21** The distribution of the radial shower position $Y$ (red line) and the radial hit position reconstructed from the beam Telescope data (blue line). The two distributions are shifted to appear one on top of the other for comparison

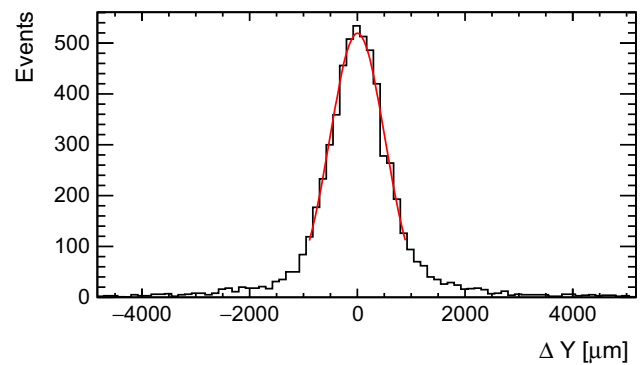

**Fig. 22** The distribution of the residuals between the reconstructed and the predicted position of the showering particle, $\Delta Y$. The curve represents the Gaussian fit to the distribution

### 4.3 Molière radius

The Molière radius, $R_\mathcal{M}$, is a characteristic constant of a material giving the scale of the transverse dimension of the fully contained electromagnetic showers initiated by an incident high energy electron or photon. By definition, it is the radius of a cylinder with axis coinciding with the shower axis, containing on average 90% of the energy deposition of the shower.

#### 4.3.1 Calculation of the Molière radius

The Molière radius, $R_\mathcal{M}$, is given by [26]

$$R_\mathcal{M} = X_0 \frac{E_s}{E_c}, \tag{3}$$

where the multiple-scattering energy $E_s = 21$ MeV, $E_c$ is the critical energy [27], and $X_0$ is the radiation length of the material.

The LumiCal prototype contains two types of absorbers made of two different tungsten alloys denoted hereafter as W-93 and W-95. The W-93 alloy contains W (93%), Ni (5.25%) and Cu (1.75%), the W-95 alloy is composed of W (95%), and the fractions of Ni and Cu are estimated as 3.5% and 1.5%, respectively. The stack contains also materials like the silicon sensors, the PCB and air. The total thickness of each material in the stack for different configurations used during the beam test are presented in Table 2.

It is important for the LumiCal operation to achieve the smalles possible transverse size of the electromagnetic shower. The shower of single high energy electrons has to be reconstructed on a widely spread background from beamstrahlung and two-photon processes. A small Molière radius facilitates this reconstruction and extends the range in the polar angle for high performance shower reconstruction [1]. In the stack used in this beam test, the main contribution to the increase of the transverse size of the shower comes from the air gap between the layers.

**Table 2** Summary of the total thickness in mm along the Z axis of the different components in the stack, for the different configurations

| Configuration | W-93 | W-95 | Air | Si | PCB | Total |
|---|---|---|---|---|---|---|
| 1 | 10.5 | 17.5 | 15.7 | 1.28 | 10.0 | 55.0 |
| 2 | 17.5 | 17.5 | 17.7 | 1.28 | 10.0 | 64.0 |
| 3 | 21.0 | 17.5 | 18.7 | 1.28 | 10.0 | 68.5 |

#### 4.3.2 The Molière radius measurement principle

For an average energy density function $F_E^V(Z, r, \varphi)$, with $V$ denoting volume, the total average energy is

$$\begin{aligned} E_{total} &= \int F_E^V(Z, r, \varphi) dV \\ &= \int F_E^V(Z, r, \varphi) dZ d\varphi r dr . \end{aligned} \tag{4}$$

Assuming that the average energy density function has a radial symmetry, the expected energy per pad tower is obtained by integration along $Z$,

$$\int F_E^V(Z, r) dZ = F_E(r) . \tag{5}$$

As noted before, on average, only 10% of the deposited energy lies outside the cylinder with a radius of one $R_\mathcal{M}$,

$$0.9 = \frac{E_{r<R_\mathcal{M}}}{E_{total}} = \frac{\int_0^{2\pi} d\varphi \int_0^{R_\mathcal{M}} F_E(r) r dr}{\int_0^{2\pi} d\varphi \int_0^\infty F_E(r) r dr} . \tag{6}$$

The LumiCal pads are long (strip like) and act like 1D integrators making it impossible to directly access the form of $F_E(r)$. Neglecting the sagitta of the pads, the energy density in the $Y$ direction can be expressed as

$$G_E(Y) = \int_{X_{min}}^{X_{max}} F_E(\sqrt{X^2 + Y^2}) dX . \tag{7}$$

Assuming a parametrized form of $F_E(r)$, its parameters can be recuperated by performing a fit to $G_E(Y)$. Depending on the form of the trial function $F_E(r)$, the integrations [Eqs. (6) and (7)] can be performed either analytically or numerically.

#### 4.3.3 Energy distribution in the transverse plane

In order to construct the average transverse distribution of the deposited energy for each $n$, the distribution $E_n$ for each event, as shown in Fig. 20, had to be shifted to the same origin. This accounts for the fact that the incoming beam had a spread. The origin was set to be the center of the middle radial pad of the instrumented area. Then the average values of $E_n$ for each $n$, $<E_n>$, were calculated. To express the fact that the pad index was shifted to a new value, the shifted pad index will be denoted by $m$. The index $m = 0$ is assigned to the central core of the shower and ranges from $-10$ to $+10$ units of pads.

The extrapolated hit position in LumiCal from the reconstructed track in the beam Telescope was used for the shifting of the data in the $Y$ direction, while in the simulation the track information was used.

The extrapolated hit position on the face of the LumiCal was also used to determine the pad (in the Y direction) in which the shower started. Only events for which the shower

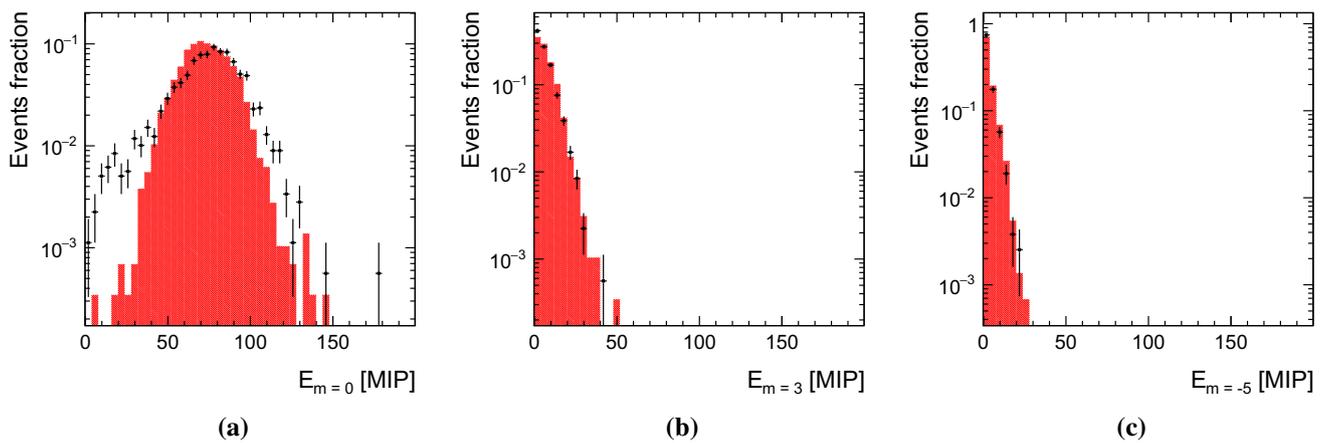

**Fig. 23** Sum of the deposited energy distribution from configuration 2, in MIP units, for the radial pads with index $m = 0, 3$ and $-5$ from the shower core at (**a**) (**b**) and (**c**), respectively. The data (dots) are compared to the MC simulation (shaded area)

started within 600 µm of the pad center were used. In addition, the difference between the extrapolated hit position in LumiCal and the position reconstructed from the LumiCal information, was required to be smaller than 2 pads (3.6 mm).

An example of the energy distributions for the shower core ($m = 0$) and for that in the wings for pads $m = 3$ and $m = -5$ are shown in Fig. 23. The data are compared to the simulation. The shape of the distribution for $m = 3$ and $m = -5$ is very well reproduced. The core distribution ($m = 0$) in the data is slightly wider and shifted, most probably due to calibration and misidentified core position. However, the mean values of the measured and simulated distributions, used for the calculation, are in agreement.

For calculating the uncertainty on the average energy of a given tower, the correlation between towers was checked [28]. Both data and MC simulations indicate that the correlation coefficients are small enough to be neglected and that energy deposits in the towers can be treated as uncorrelated. Thus the uncertainty of each tower consists of the statistical uncertainty and the 5% systematic uncertainty from the energy calibration.

### 4.3.4 Functional form

The distribution of the average deposited energy in the transverse plane is symmetric with respect to the longitudinal shower axis and does not depend on the azimuthal angle. Its radial dependence is characterized by a narrow core and a broad base. The function used to describe the average transverse energy profile of the shower is a Gaussian for the core part and a form inspired by the Grindhammer-Peters parametrisation [29,30] to account for the tails,

$$F_E(r) = A_C e^{-\left(\frac{r}{R_C}\right)^2} + A_T \frac{2r^\alpha R_T^2}{(r^2 + R_T^2)^2}, \quad (8)$$

and integrating over the horizontal position $X$, the vertical energy distribution $G_E(Y)$ is

$$G_E(Y) = \int_{X_{min}}^{X_{max}} A_C e^{-\left(\frac{\sqrt{X^2+Y^2}}{R_C}\right)^2}$$
$$+ A_T \frac{2\left(\sqrt{X^2+Y^2}\right)^\alpha R_T^2}{((X^2+Y^2)+R_T^2)^2} dX . \quad (9)$$

Here $A_C$, $R_C$, $A_T$, $R_T$ and $\alpha$ are parameters to be determined by fitting the function to the measured distribution. The range $(X_{min}, X_{max})$ is defined by the sensor geometry which corresponds to two sectors. The integration is performed numerically. By fitting $G_E(Y)$ to the shower transverse profile, and finding its parameters, the original $F_E(r)$ can be recovered and then used to determine $R_\mathcal{M}$ from Eq. (6).

The form of the function (8) was chosen because it describes best the data, though it is known [26] that such functions can be used to describe the data only up to 3.5 Molière radii (99% of the energy containment).

### 4.3.5 Simulation of a complete Calorimeter

For the numerical integration, the integration limits of the normalization integral in the denominator of Eq. (6) must be chosen such that the relevant range in $r$ is covered. The part of the function (8) which describes the energy deposition far from the shower axis does not provide a fast convergence of that integral. That is why the solution of the Eq. (6) may depend on the upper limit of the integration.

To this end, an additional MC simulation with a slightly modified geometry of the calorimeter was used in order to estimate the systematic uncertainties caused by the choice of the trial function (8) and its normalization in (6). For this study, the sensitive detector of the sampling calorimeter was implemented with a fine granularity of $0.5 \times 0.5$ mm$^2$ and the transverse size of the calorimeter was extended up to 40 × 40 cm$^2$. Twenty thousand electrons with 5 GeV momenta were simulated.

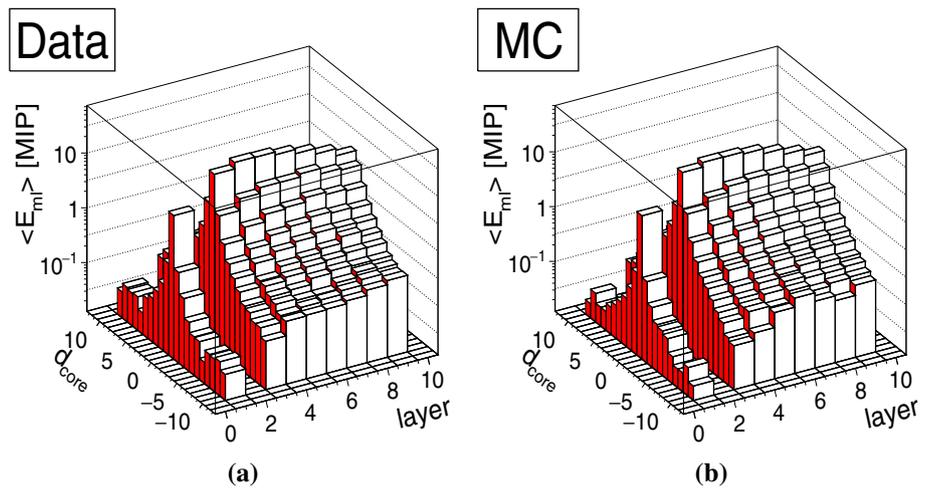

**Fig. 24** A lego plot of the transverse profile $<E_m>$, as a function of $d_{core}$ in units of pads, for each layer from the beam-test data (**a**) and for the MC simulation (**b**)

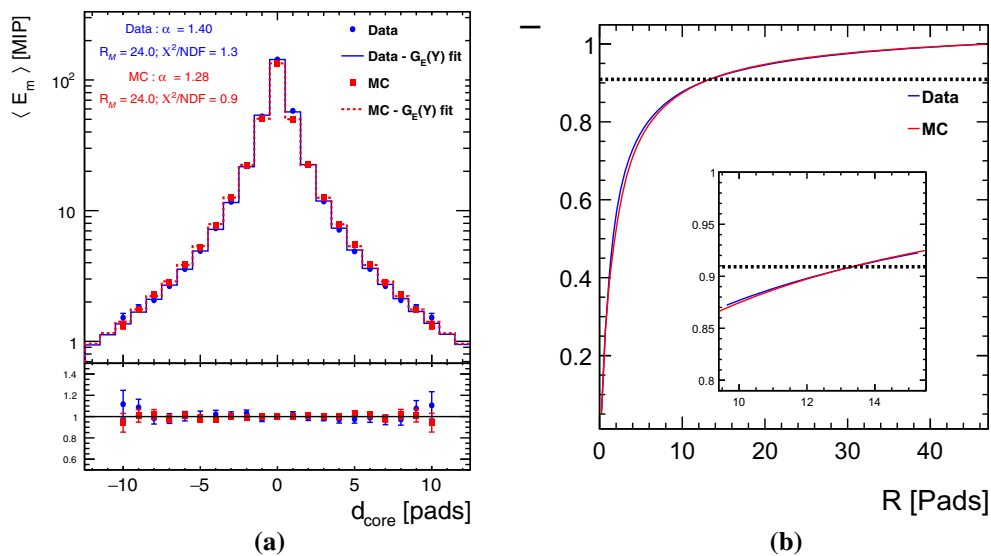

**Fig. 25** **a** The shower transverse profile $<E_m>$, as a function of $d_{core}$ in units of pads, of the joint distribution of all three configuration from beam-test data and the MC simulation, after symmetry corrections and fit. The lower part of the figure shows the ratio of the distributions to the fitted function, for the data (blue) and the MC (red). **b** The integral on $F_E(r)$, $I$, that was extracted from the fit in (**b**). The insert in (**a**) shows an expanded view of the region $10 < R < 15$ pads

In this detailed simulation, 99% of the deposited energy for the second configuration of the beam test setup was contained inside an area limited by a radius of R = 84.5 ± 0.5 mm. This limit was used for the integration of (6) and the right part of the equation was changed accordingly to 0.9091. This simulation also allowed to estimate the fraction of the energy collected by the part of the LumiCal sensor which was connected to the readout electronics (97.8 %) and use this fraction as an additional condition to constrain the trial function in the fit. Finally it was established that the effect of the approximation of the circular shape of LumiCal sensor pads by strips is negligibly small in the center of the distribution (9) and it is below 2% in the tails where the statistical uncertainty is significantly higher.

### 4.3.6 Molière radius results

The three configurations, when properly combined, allow to follow the development of the shower in more detail than each configuration separately and in steps of 1 $X_0$. The average energy deposition in each radial distance from the shower core per layer is denoted by $<E_{ml}>$. For the layers that are probed more than once, the appropriate average is used. The variable $<E_{ml}>$ as a function of the distance from the shower core, $d_{core}$, is plotted for each layer in the lego plot presented in Fig. 24 for data and for MC. The simulation is in good agreement with the measurements.

**Table 3** The Molière radius results, data and MC simulation, together with the $\chi^2/NDF$ and the value of $\alpha$ from the fit to the radial distribution

| Data | | | MC | | |
|---|---|---|---|---|---|
| $R_\mathcal{M}$ [mm] | $\chi^2/NDF$ | $\alpha$ | $R_\mathcal{M}$ [mm] | $\chi^2/NDF$ | $\alpha$ |
| $24.0 \pm 0.6$ | 1.3 | $1.40 \pm 0.03$ | $24.0 \pm 0.6$ | 0.9 | $1.28 \pm 0.03$ |

In order to build the shower transverse profile for all measured layers, the average energy deposited per pad, $<E_m>$, is constructed as the sum of $<E_{ml}>$ over all layers. The shower transverse profile, expressed as the distance from the shower core, $d_{core}$ in units of pads, is presented in Fig. 25. The fit and the solution of the algebraic Eq. (6) were calculated numerically using the ROOT package. For the fitting procedure, the integral of the function inside the bin, normalized by the bin width, was used.

The results of the calculation are summarized in Table 3, where the Molière radius is given for the combination of all configurations used in the beam test and for the corresponding MC simulation. In order to estimate the systematic uncertainty of the numerical calculation, the fit to the shower transverse profile and the Molière radius extraction was repeated 1000 times for each set. In each repetition every point in the transverse profile was shifted randomly according to a Gaussian distribution with a $\sigma$ equal to the point uncertainty, before the fit procedure. The simulation results are in very good agreement with that of the data.

It is worth noting that the value obtained by using the formula for composite material given in [26] is lower ($\sim 17$ mm) than obained from this analysis. This is due to the fact that the composite material formula is not precise enough to describe a layer structure with the thickness of the layers large compared to the appropriate radiation length [31], as was the case in the present setup.

The systematic uncertainty was estimated by taking into account the uncertainty of the normalization factor used from the complete calorimeter, from the integration range in $X$ and of the constraint of 97.8%, also from the complete calorimeter. Adding them in quadrature, the total systematic uncertainty results in $\pm 1.5$ mm. Thus the result of the Molière radius determination is

$$R_\mathcal{M} = 24.0 \pm 0.6(\text{stat.}) \pm 1.5(\text{syst.})\text{mm} . \qquad (10)$$

The MC simulation is in very good agreement with the measured value.

## 5 Conclusions

For the first time a multi-plane operation of a prototype of a luminometer designed for a future $e^+e^-$ collider detector was carried out. The test was performed in the CERN PS accelerator T9 beam line with a 5 GeV beam. The development of the electromagnetic shower was investigated and shown to be well described by a GEANT4 Monte Carlo simulation. The shower position resolution for 5 GeV electrons was measured to be $505 \pm 10$ µm and the effective Molière radius of the configurations used in this beam test was determined to be $24.0 \pm 0.6$ (stat.) $\pm 1.5$ (syst.) mm.

The paper demonstrates that major components for a luminometer to be used at a future experiment at CLIC or ILC, developed by the FCAL collaboration, can be operated as a system. The performance in reconstructing electromagnetic showers is well reproduced by Monte Carlo simulations.

**Acknowledgements** This study was partly supported by the Israel Science Foundation (ISF), Israel German Foundation (GIF), the I-CORE program of the Israel Planning and Budgeting Committee, Israel Academy of Sciences and Humanities, by the National Commission for Scientific and Technological Research (CONICYT - Chile) under Grant FONDECYT 1170345, by the Polish Ministry of Science and Higher Education under Contract Nrs 3585/H2020/2016/2 and 3501/H2020/2016/2, the Rumanian UEFISCDI Agency Under Contracts PN-II-PT-PCCA-2013-4-0967 and PN-II-ID-PCE-2011-3-0978, by the Ministry of Education, Science and Technological Development of the Republic of Serbia within the project Ol171012, and by the European Union Horizon 2020 Research and Innovation programme under Grant Agreement No. 654168 (AIDA-2020).

## References

1. H. Abramowicz et al., [FCAL Collaboration], Forward Instrumentation for ILC Detectors. JINST **5**, 12002 (2010)
2. T. Behnke, J.E. Brau, B. Foster, J. Fuster, M. Harrison, J.M. Paterson, M. Peskin, M. Stanitzki, N. Walker, H. Yamamoto (eds) The International Linear Collider Technical Design Report - Volume 1: Executive Summary. arXiv:1306.6327
3. L. Linssen, A. Miyamoto, M. Stanitzki, H. Weerts (eds) CLIC CDR, Physics and Detectors at CLIC. CERN-2012-003 ; ANL-HEP-TR-12-01 ; DESY-12-008 ; KEK-Report-2011-7. arXiv:1202.5940 [physics.ins-det]
4. Ch. Grah, A. Sapronov, Beam parameter determination using beamstrahlung photons and incoherent pairs. JINST **3**, 10004 (2008)

5. P. Bambade, V. Drugakov, W. Lohmann, The impact of BeamCal performance at different ILC beam parameters and crossing angles on stau searches. Pramana **69**, 1123 (2007)
6. T. Behnke, J.E. Brau, P. Burrows, J. Fuster, M. Peskin, M. Stanitzki, Y. Sugimoto, S. Yamada, H. Yamamoto (eds), The international linear collider technical design report-volume 4: detectors. arXiv:1306.6329
7. H. Abramowicz et al., [FCAL Collaboration], Performance of fully instrumented detector planes of the forward calorimeter of a Linear Collider detector. JINST **10**, P05009 (2015)
8. F.-X. Nuiry, Collected documents on the FCAL-AIDA precision mechanical infrastructure and tungsten plates. https://edms.cern.ch/document/1475879/
9. M. Idzik, S. Kulis, D. Przyborowski, Development of front-end electronics for the luminoisty detector at ILC. Nucl. Instr. Methods A **608**, 169 (2009)
10. M. Idzik, K. Swientek, T. Fiutowski, S. Kulis, D. Przyborowski, A 10-bit multichannel digitizer ASIC for detectors in particle physics experiments. IEEE Trans. Nucl. Sci. **59**, 294 (2012)
11. S. Kulis, A. Matoga, M. Idzik, K. Swientek, T. Fiutowski, D. Przyborowski, A general purpose multichannel readout system for radiation detectors. JINST **7**, T01004 (2012)
12. U.I. Uggerhoj, T.N. Wistisen, Detector tests and readiness for 2014, CERN NA63. CERN-SPCS-2013/036 / SPSC-SR-127 (2013)
13. J. Baudot et al., First test results of MIMOSA-26: a fast CMOS sensor with integrated zero suppression and digitized output. http://inspirehep.net/record/842163?ln=en
14. D. Cussans, A trigger/timing logic unit for ILC Test-beams. In: Proceedings of the Topical Workshop on Electronics for Particle Physics, TWEPP-07, Prague, Czech Republic, 3–7 September 2007
15. D. Cussans, Description of the JRA1 trigger logic unit (TLU). EUDET-Memo-2009-4. https://www.eudet.org/e26/e28/e42441/e57298/EUDET-MEMO-2009-04.pdf
16. J. Baudot, TAF short manual. http://www.iphc.cnrs.fr/IMG/pdf/taf_shortdoc.pdf, http://www.iphc.cnrs.fr/TAF.html
17. O. Rosenblat, Uniformity of detector prototypes for instrumentation in the very forward region of future linear colliders. MSc Thesis, Tel Aviv University (2016)
18. T.L. Bienz, Strangeonium spectroscopy at 11 GeV/c and Cherenkov Ring Imaging at the SLD. PhD dissertation, Stanford University (1990)
19. J. Moron, Development of novel low-power, submicron CMOS technology based, readout system for luminosity detector in future linear collider. Ph.D. Thesis, AGH - UST, Cracow (2015)
20. J. Aguilar, Luminosity detector at ILC: Monte Carlo simulations and analysis of test beam data. MSc Thesis, WFiIS AGH and IFJ-PAN, Krakow (2012)
21. S. Agostinelli et al., GEANT4—a simulation toolkit. Nucl. Inst. Methods A **506**, 250 (2003)
22. R. Brun, F, Rademakers, ROOT—an object oriented data analysis framework. Nucl. Inst. Methods A **389**, 81 (1997). See also http://root.cern.ch/
23. P. Mora de Freitas, H. Videau, Detector simulation with Mokka and Geant4: Present and future. Technical Report LC-TOOL- 2003-010, LC Note, March 2003. Mokka web site. http://mokka.in2p3.fr
24. QGSP_BERT, Reference physics lists. http://geant4.cern.ch/support/proc_mod_catalog/physics_lists/referencePL.shtml
25. E. Longo, I. Sestili, Monte Carlo calculation of photon-initiated electromagnetic showers in lead glass. Nucl. Inst. Methods **128**, 283 (1975)
26. K.A. Olive et al., Particle Data Group, Chin. Phys. C **38**, 090001 (2014). and 2015 update
27. B. Rossi, K. Greisen, Cosmic-ray theory. Rev. Mod. Phys. **13**, 240 (1941)
28. I. Levy, Forward Calorimetry for future linear colliders. PhD Thesis, Tel Aviv University, Tel Aviv (2017)
29. G. Grindhammer et al., In: Proceedings of the Workshop on Calorimetry for the Supercollider, Tuscaloosa, AL, March 13-17, 1989, edited by R. Donaldson and M.G.D. Gilchriese (World Scientific, Teaneck, NJ, p. 151) (1989)
30. G. Grindhammer, M. Rudowicz, S. Peters, The Parameterized Simulation of Electromagnetic Showers in Homogeneous and Sampling Calorimeters. arXiv:hep-ex/0001020
31. R. Wigmans, *Calorimetry—Energy measurement in Particle Physics* (Oxfort University Press, Oxfort, 2017)